\documentclass[prl,twocolumn,amssymb,floatfix]{revtex4}

\usepackage{graphicx}
\usepackage{color}
\usepackage{epsfig}
\usepackage{latexsym}
\usepackage{bm}
\usepackage{subfig}
\usepackage{mathtools}
\usepackage{eucal}
\usepackage{caption}
\usepackage{ragged2e}
\usepackage[colorlinks=true, allcolors=blue]{hyperref}

\begin{document}

\renewcommand{\ni}{{\noindent}}
\newcommand{\dprime}{{\prime\prime}}
\newcommand{\be}{\begin{equation}}
\newcommand{\ee}{\end{equation}}
\newcommand{\bea}{\begin{eqnarray}}
\newcommand{\eea}{\end{eqnarray}}
\newcommand{\nn}{\nonumber}
\newcommand{\la}{\langle}
\newcommand{\ra}{\rangle}
\newcommand{\dg}{\dagger}
\newcommand{\pll}{\parallel}
\newcommand{\sumr}{\sum_{\vr}}
\newcommand{\ua}{\uparrow}
\newcommand{\da}{\downarrow}
\newcommand{\oi}{{\Omega_{I}}}
\newcommand{\oo}{{\Omega}}
\newcommand{\kon}{{k_{\text{on}}}}
\newcommand{\kof}{{k_{\text{off}}}}
\newcommand{\logn}{{\text{ln}}}
\newcommand{\Phim}{{{\Phi_0}}}
\newcommand{\Psim}{{{\Psi_0}}}
\newcommand{\oom}{{{\Omega_0}}}
\newcommand{\Phit}{{\tilde{\Phi}}}
\newcommand{\Psit}{{\tilde{\Psi}}}
\newcommand{\oomt}{{\tilde{\Omega}}}
\newcommand{\ep}{{\mathcal{E}}_p}
\newcommand{\oot}{{\Omega_{\text{Total}}}}

\title{Non-equilibrium structure and relaxation in active microemulsions}

\author{{\normalsize{}Rakesh Chatterjee$^{1,2}$}
{\normalsize{}}}

\author{{\normalsize{}Hui-Shun Kuan$^{1,2,3}$}
{\normalsize{}}}

\author{{\normalsize{}Frank J\"ulicher$^{4,5,6}$}
{\normalsize{}}}

\author{{\normalsize{}Vasily Zaburdaev$^{1,2}$}
{\normalsize{}}}

\affiliation{$^1$ Department of Biology, Friedrich-Alexander-Universit\"at Erlangen-N\"urnberg, Erlangen, Germany \\ $^2$ Max-Planck-Zentrum f\"ur Physik und Medizin, Erlangen, Germany\\$^3$ TSMC, Hsinchu Science Park, Hsinchu, Taiwan
\\$^4$ Max Planck Institute for the Physics of Complex Systems, Dresden, Germany\\$^5$
Center for Systems Biology Dresden, Dresden, Germany\\$^6$
Cluster of Excellence Physics of Life, TU Dresden, Dresden, Germany}

\begin{abstract}

Microphase separation is common in active biological systems as exemplified by the separation of RNA and DNA-rich phases in the cell nucleus driven by the transcriptional activity of polymerase enzymes acting similarly to amphiphiles in a microemulsion. Here we propose an analytically tractable model of an active microemulsion to investigate how the activity affects its structure and relaxation dynamics. Continuum theory derived from a lattice model exhibits two distinct regimes of the relaxation dynamics and is linked to the broken detailed balance due to intermittent activity of the amphiphiles.   

\end{abstract}

\maketitle
A broad range of complex biological phenomena \cite{sartoria,tailleur_nature,marini,fodor} as well as man-made systems of active matter \cite{sriram_rev,h-su,katuri} motivated the development of theoretical approaches either generalized from equilibrium statistical physics or fully non-equilibrium.  In some regimes, equilibrium concepts remain useful
even in an intrinsically out of equilibrium contexts such as a living cell \cite{kreysing_frank}, sometimes it is applied phenomenologically \cite{prost_2015,transtrum}, and sometimes essentially non-equilibrium models are put forward \cite{huishun_prl}. Further examples include the hydrodynamics of flocks and active self-propelled particles \cite{toner,marchetti1}, and the active gel theory of the cell skeleton and beyond \cite{joanny1}. Currently, a broad range of biological applications is found in the theory of liquid-liquid phase separation \cite{liq-liq1,liq-liq2} and specifically in the concept of active microemulsion \cite{cates_08,cates_12,cates_18,thutu,frank_emulsion,frey_22,frey_24,Lemma, Carrere,Fausti,Gulati} that describes the organization of complex multi-phase systems. However, theoretical understanding of nonequilibrium effects in microemulsion systems is still very limited. 
Here, motivated by a biological process of chromatin organization by transcription, we formulate an active microemulsion model and investigate the effects of activity on its structure and relaxation dynamics.

Transcriptionally active nuclei exhibit the phenomenon of microphase separation into DNA-rich and RNA-rich domains \cite{lennart_1}. It has been suggested that polymerase molecules engaged in the process of transcription can have an effect similar to amphiphiles in a prototypical oil-and-water emulsion \cite{me_1,me_2}. The process of transcription itself is active as manifested by its intermittent dynamics and by the production of the RNA-component. While it is well understood how in passive systems the presence of amphiphiles determines the process of microphase separation or how intermittent attractive interactions may drive microphase separation (via several theoretical and numerical approaches \cite{pheno_1,pheno_2,pheno_3,pheno_4,pheno_5,latt_1,latt_2,latt_3,latt_4,
latt_7,grosberg_pre,golestanian,grosberg_prl,alston}), how the intermittent activity of amphiphiles could change the microemulsion structure and relaxation dynamics has not yet been explored. 

Starting with a generalization of the Gompper-Schick lattice model of a microemulsion \cite{gompper_prl} we show that an amphiphile that intermittently switches from an amphiphilic to an inert state can change the structure of the microemulsion. We then coarse-grain the lattice model to develop a continuum version of the theory and find a two-regime relaxation dynamics of active microemulsions. The two regimes of relaxation can be attributed to the fast amphiphile turnover and slower diffusion of phases respectively. Finally, we can link the local entropy production at the amphiphile-rich interface to the nonequilibrium amphiphile turnover dynamics. Our theory will help to better understand the non-equilibrium patterning processes in active multi-phase systems.

\textit{Biological motivation and model.-}
To investigate the role of intermittent amphiphilic activity, we establish a phenomenological model, which qualitatively captures the experimental observations on microphase separation in the cell nucleus \cite{lennart_1}.
DNA and mRNA, produced during transcription, are the two phase-separating components. Polymerase molecules that attach to DNA and perform transcription by producing mRNA-transcripts thereby connect the two segregating phases and therefore act as amphiphiles. This configuration with three basic components is similar to the ternary oil-water-amphiphile system, which exhibits two- and three-phase coexistence \cite{latt_4}. However, unlike in the classical example, amphiphile-like activity of polymerases is intermittent \cite{schneider,wagh,wang}: polymerases bind to DNA and engage in transcription, moving persistently along the DNA strand but then unbind from DNA when reaching the end of the transcribed gene. We thus aim to model this intermittent amphiphile dynamics by introducing the switching of the amphiphile between the state when it exerts its amphiphilic property and reduces the system's energy at the interface between the two phases, and ``inert'' when it does not. In the cell nucleus, RNA is continuously produced and exported to the cytoplasm for translation, maintaining a fairly constant RNA/DNA ratio in a homeostatic state \cite{lennart_1, berry}. For instance, in the context of embryonic development, the variation in some of the accumulated mRNAs species was reported to be $\sim 8\%$ \cite{little}. We therefore assume that the composition of our system is not changing over time.

\begin{figure}[ht]
\includegraphics[width=8cm,height=4.7cm]{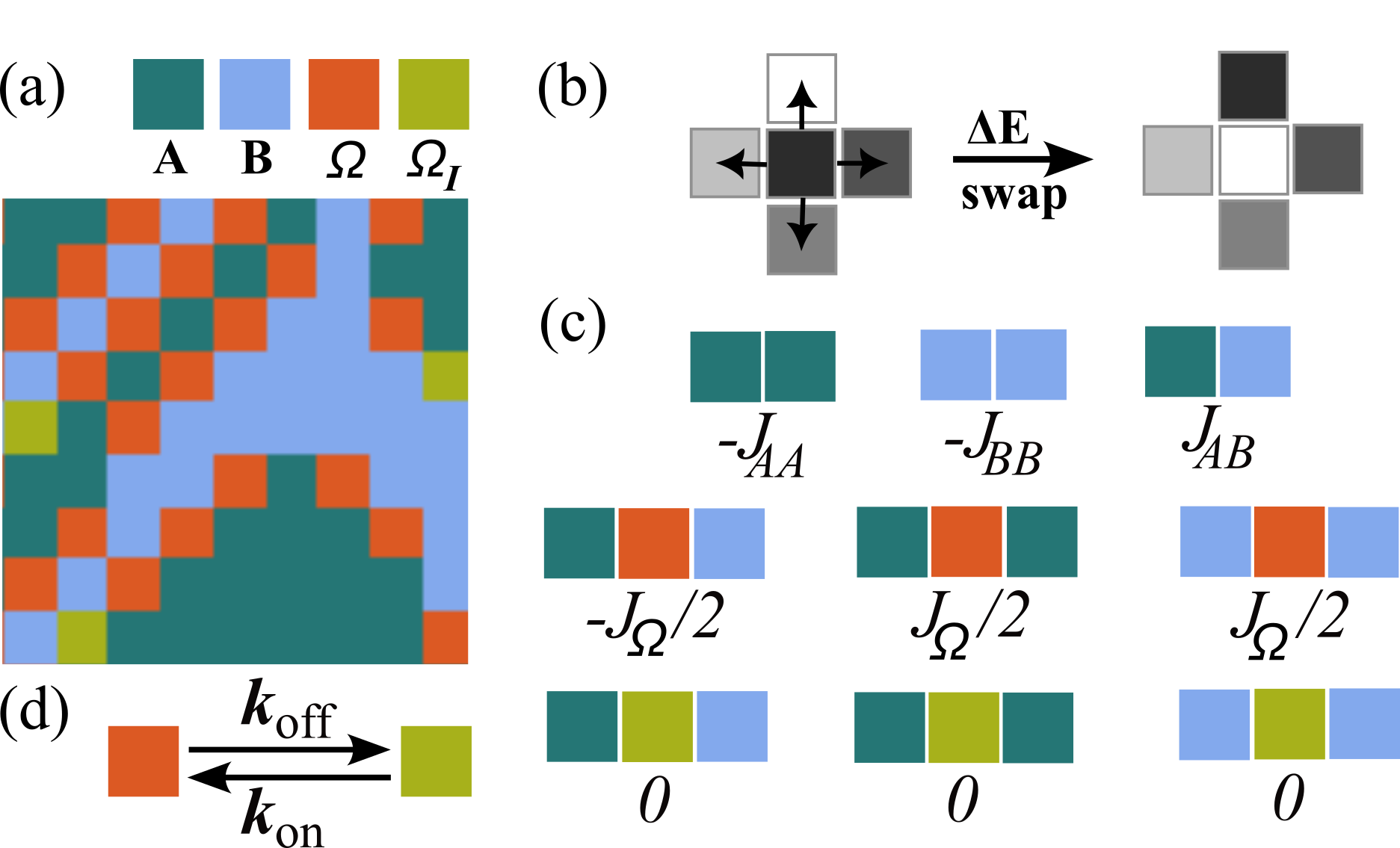}
\caption{Schematic representation of the lattice model of an active microemulsion. (a) Lattice species. Each lattice site is occupied by any one of the components $A$ (teal), $B$ (blue) and particles that are amphiphilic $\oo$ (orange) or inert $\oi$ (olive).  (b) Lattice dynamics. For a given site, one of its four nearest neighbors is randomly selected. Their swap is attempted based on the system's energy difference before and after the swap $\Delta E$. (c) Contributions to system's energy. Particles of the same components attract ($-J_{AA}$, $-J_{BB}$), and there is a repulsion between the two different components ($J_{AB}$). Particles in the amphiphilic state $\oo$ reduce system's energy by $J_{\oo}/2$ when flanked by $A$ and $B,$ an increase it by $J_{\oo}/2$ when flanked by the same component.  Inert particles do not contribute to system's energy. (d) Intermittent amphiphile. In active microemulsion model,  intermittency of amphiphiles is introduced by switching $\oo$ to $\oi$ with rate $\kof$ and the reverse with rate $\kon.$ }
\label{fig:model}
\end{figure}

\textit{Lattice model of microemulsion.-}
We first start with the lattice-based implementation of the active microemulsion model building upon the work of Gompper and Schick
\cite{gompper_prl,gompper_prb,gompper_pra}. A square lattice of size $L\times L$ contains two phase-separating components $A$ and $B$, while $\oo$ acts as an amphiphile, with an individual element of each component occupying a single lattice site (see Fig.\ref{fig:model}). 
We use the size of the single lattice site as the length scale and set it to be equal to $1$ for the rest of the manuscript.  $A$ and $B$ exhibit self-affinity with parameters $J_{AA}$ and $J_{BB}$ respectively and repel each other as parameterized by $J_{AB}$. 
Here, for simplicity, we will consider $J_{AA}=J_{BB}$ and no repulsion $J_{AB}=0$. This choice is aimed to mimic  the tendency of RNA and RNA-binding proteins to create condensates (in vivo and in vitro) which segregate from DNA-rich areas leading to their phase separation, cf.\cite{lennart_1}.  Particles $\oo$ in the ``amphiphilic'' state  reduce the system's energy by the value $J_{\oo}/2,$ denoted as amphiphile-mediated interaction, if placed on a line flanked by $A$ and $B$ species.  Energy is increased by $J_{\oo}/2$ if the amphiphile is flanked by the particles of the same type, $A$ or $B$ (see Fig.\ref{fig:model}). 
Amphiphiles in the inert state $\oo_I$ do not contribute to the energy of the system. Here we introduce intermittent activity by switching the functional amphiphiles to the inert state with rate $\kof$ and back to the amphiphilic state with rate $\kon,$ with the total amphiphile particle content conserved $\oo_{\text{Total}}= \oo+ \oo_I$.

The Hamiltonian of a four-component system (two phases and two states of the amphiphile) is described in terms of occupancy operators $\eta_i^{A,B,\oo,\oi}$, which are equal to unity if a site $i$ is occupied by the respective component $A$, $B$, $\oo$, $\oi$ and zero otherwise, and interaction strengths $J_{ij}$ for two phases $i,j=\{A,B\}$ and amphiphile-mediated interaction $J_{\Omega}$:

\small
\bea
{\cal{H}}\!\! &=&\!\! -\frac{1}{2}\sum_{<ij>} \Big(J_{AA} \eta_i^A \eta_j^A + J_{BB}\eta_i^B\eta_j^B + J_{AB}(\eta_i^A\eta_j^B + \eta_i^B\eta_j^A)\Big) \nn \\
&-&\!\! \frac{J_{\oo}}{2}\sum_{<ijk>} \Big(\eta_i^A\eta_j^{\oo}\eta_k^B + \eta_i^B\eta_j^{\oo}\eta_k^A - \eta_i^A\eta_j^{\oo}\eta_k^A - \eta_i^B\eta_j^{\oo}\eta_k^B \Big)
\label{h_eqn}
\eea
\normalsize
The first term is the sum over same and distinct component pair interactions. The second term represents the contribution of amphiphile-mediated interaction.
There is an intermittent dynamics of amphiphile activity $\eta_i^{\oo}  \xrightarrow{\kof}  \eta_i^{\oi} ~,~
\eta_i^{\oi} \xrightarrow{\kon}  \eta_i^{\oo}$, and the amphiphiles in the inert state $\oi$ do not contribute to the Hamiltonian.

We use kinetic Monte-Carlo simulations to investigate the lattice model. The system evolves from a random initial configuration through the pair-swap dynamics of the two nearest neighbors randomly chosen within $4$-neighborhood. The swap is executed with probability $1$ if the system's energy difference $\Delta E$ calculated from Eq. (\ref{h_eqn}) after the proposed swap is negative, and with probability
$P_{\text{swap}}=\text{exp}(-\Delta E/k_BT)$ otherwise. All interaction energies are normalized by $k_BT$, where $k_B$ is the Boltzmann constant and $T$ is the temperature. To provide a fair comparison of an active and a passive system, in the passive counterpart, we also introduce two particles species (with amphiphilic activity and inert) which are present in fixed proportions ($\kon/(\kon+\kof)$ and $\kof/(\kon+\kof)$ respectively) and never switch their identities.

\begin{figure}[ht]
\includegraphics[width=8.2cm,height=2.6cm]{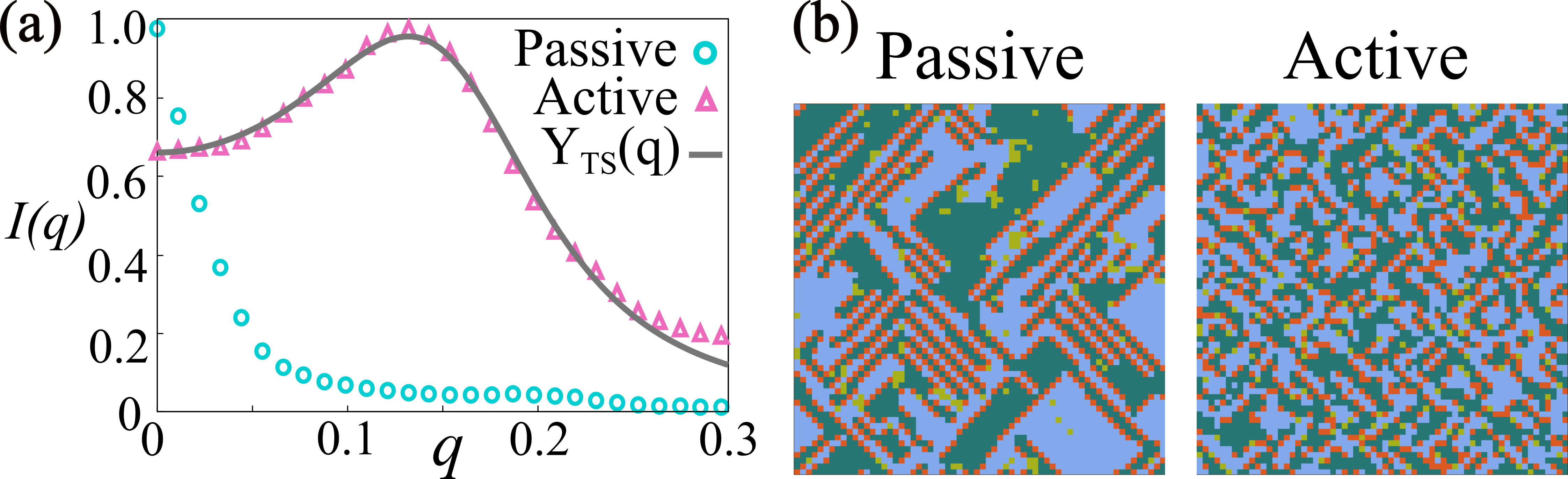}
\caption{Structure factor and lattice configurations for passive and active systems.  (a) Structure factor as a function of the magnitude of the wave vector $q$ of lattice configurations in passive and active systems.  Unlike the disordered unstructured phase in the passive case (blue circles), the disordered structured or microemulsion phase in the active case (magenta triangles) has a peak at $q>0$. The form of the intensity distribution in the active case is compared with the phenomenological Landau theory of scattering intensity distribution of a microemulsion ($Y_{TS}(q)$) which shows a $q^{-4}$ decay for large $q$ values (see text). (b) Corresponding lattice configurations of passive and active cases display how amphiphile turnover transforms the structure of the disordered phase.  ($A=B=0.35$, $\oot=0.30,$ $k_{\text{on}}=0.9$, $k_{\text{off}}=0.1$, $J_{AA}=J_{BB}=2.5, ~J_{AB}=0$ and $J_{\oo}=3.0$, $L=64$.) }
\label{fig:str_factor}
\end{figure}
For the passive system, the two- and three-phase coexistence, as well as phase transitions between ordered lamellar phase to disordered structured (microemulsion) phase, (both characterized by the presence of a peak in their structure factor function), and to disordered structure-less phases (no peak) were previously described by this model \cite{gompper_prb}.  The phase diagram for ternary systems was examined through the mean-field theory. The disordered structured phase occurs at higher concentrations and temperatures, while the structure-less phase emerges at low concentrations and high temperatures. As temperature decreases, at low concentrations, the structure-less phase transitions into an oil-water coexistence state, while at higher concentrations, the system shifts to a lamellar phase with organized layers. Our primary interest lies in the subregion of this phase diagram corresponding to the disordered microemulsion phase.  Here we focus on the effects of amphiphiles' turnover on the system's structure and dynamics.

To characterize the organization of microemulsion we measured the static structure factor which is defined as the Fourier transform of the density-density correlation function of any of the components $A$, $B$ and $\oo$: $I(q)=\langle \rho_{_{A}}(q)\rho_{_{A}}(-q) \rangle_q$ \cite{isf_def}. Here, $\rho_{_{A}}(q)$ is the Fourier transform of the image intensity and $q=|\mathbf{q}|=\sqrt{q_x^2+q_y^2}$ is the absolute value of the wave vector. We collect the images of the steady-state lattice configurations (after a prolonged equilibration phase,  see Fig.S1).   
In Fig.\ref{fig:str_factor}a,  we show $I(q)$ thus obtained from lattice simulations. 
For the selected parameters, we see that a passive system corresponds to a disordered unstructured phase as the structure factor has no peak (see open circles in Fig.\ref{fig:str_factor}a and the snapshot of lattice configuration, panel b, left). Remarkably, the same composition with amphiphile turnover is a disordered structured phase (microemulsion) phase, as seen by the occurrence of a distinct peak in the structure factor (shown by open triangles in Fig.\ref{fig:str_factor}a and respective lattice snapshot in panel b, right). Interestingly, we can use the expression of the structure factor $Y_{TS}(q)$ derived from a phenomenological equilibrium microemulsion theory \cite{teubner} as a fitting function for the case of active microemulsion
$Y_{TS}(q) \sim (a+c_1 q^2 + c_2 q^4)^{-1},$ 
which with negative $c_1$ reaches a maximum before it decays as $q^{-4}$ (see additional fits for passive and active cases in Fig.S2).

Thus, we see that the activity due to intermittent dynamics of amphiphiles can significantly change the structural organization of the microemulsion without altering its composition.  We also notice that the dynamics of two systems (see Supplemental Movies 1 and 2) is also distinctly different. To get a better insight into dynamics, we can take advantage of the lattice model that allows for coarse-graining and the derivation of a continuum theory.

\begin{figure}[ht]
\includegraphics[width=8.4cm,height=5cm]{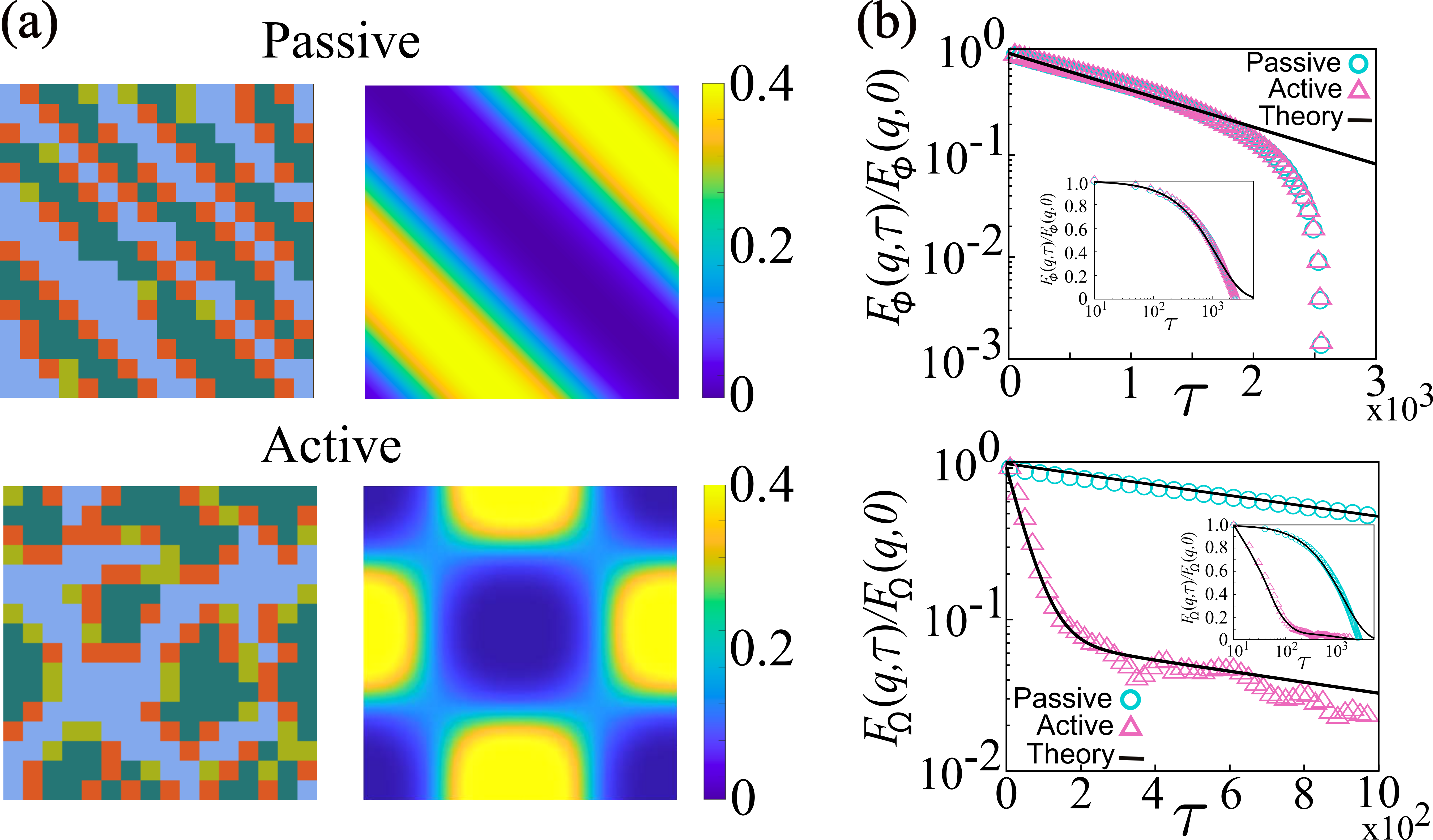}
\caption{Continuum theory and relaxation dynamics. (a) Comparison of lattice and continuum models' simulations. Lattice configuration and the corresponding density map from continuum theory show a lamellar-like structure in the passive case and a microemulsion structure in the active case ($A = B = 0.30,$ $\oot = 0.40,$ $J_{AA} = J_{BB} = 1.5,$ $J_{\Omega} = 4.0$, $\kon = 0.75,$ $\kof = 0.25.$) (b) Intermediate scattering function (ISF) as a function of the delay time $\tau$.  ISF is calculated for the magnitude of the wave vector $q=\sqrt{2}/L$ 
for the modified density component $\Phi$ (top) and $\oo$ (bottom), parameters are the same as in Fig. 2. The plots show the same exponential relaxation dynamics for $\Phi$ and $\oo$ in the passive case. In the active case, relaxation of $\oo$ is different and described as a superposition of two exponentials.  Points are from lattice simulations and solid lines are from the coarse-grained continuum theory. Insets shows the same plots in the log-linear scale. Time scale is given in Monte Carlo steps with the parameters of the continuum theory adjusted to match this time.}
\label{fig:relaxation}
\end{figure}

\textit{Coarse grained continuum theory.-}
Considering the nearest neighbor interactions we coarse grain the system to derive the continuum description for the free energy density as discussed in Supplemental Material \cite{supp}:
\bea
{\cal{F}} &=&-\frac{J_{AA}}{2}\Big(A^2 - \frac{1}{2}(\nabla A)^2\Big)-\frac{J_{BB}}{2}\Big(B^2 - \frac{1}{2}(\nabla B)^2\Big) \nn \\
 &-& J_{\oo}\Big(\oo (2AB-4 \nabla A \cdot \nabla B)) - \oo(A^2+B^2-2(\nabla A)^2 \nn \\
&-& 2(\nabla B)^2\Big) +A\logn A + B\logn B + \oo \logn \oo + \oi \logn \oi,
\label{free_en_cont1}
\eea
where $A$, $B$ and $\oo$ denote the densities of the corresponding components (and $\oi=1-A-B-\oo$).
Next, we can derive the dynamical equations for different phases. The particle number conservation of $A$, $B$ and $\oo$ can be expressed by the continuity equations.
In the linear response regime, the particle current is proportional to the thermodynamic force of the gradient of the chemical potential and thus can be written in terms of the free energy functional:
\bea
\partial_t A &=& \Gamma_{_{A}} \nabla \cdot \Big( \nabla \frac{\delta {\cal{F}}}{\delta A}\Big), \quad
\partial_t B = \Gamma_{_{B}} \nabla \cdot \Big( \nabla \frac{\delta {\cal{F}}}{\delta B}\Big), \\
\partial_t \oo &=& \Gamma_{_{\oo}} \nabla \cdot \Big( \nabla \frac{\delta {\cal{F}}}{\delta \oo}\Big) + \kon \oi - \kof \oo,
\label{eqn_dynamical_form}
\eea
where $\Gamma_{_{A,B,\oo}}$ is the diffusion coefficient for particles at unit temperature. For constant temperature, and assuming that all particles are moving through nearest neighbor swap at the same rate, the diffusion coefficients for all types of particles are considered to be the same. The nonequilibrium contribution of the amphiphile turnover is in the chemical reaction turnover terms in the last equation, which is not derived as a part of the free-energy current. We now perform numerical simulations of this continuum model and show that it qualitatively reproduces the phase separation patterns as previously observed in the lattice simulations (see Fig. \ref{fig:relaxation}a), where parameters are such that the passive system is in the lamellar state, and its active counterpart is in the microemulsion state. We next aim to calculate the intermediate scattering function (ISF) \cite{isf_3} to quantify the relaxation dynamics in our system. 

To simplify calculations further, we introduce a change of variables by two modified local densities $\Phi=A+B$ and $\Psi=A-B$ which will enable us to perform a linear stability analysis. We further assume that density of $A$ and $B$ phases are equal as well as the pair interactions of similar kind of particles $(J_{AA}=J_{BB}).$ 
The diffusion coefficient for the modified local densities turn out to be $\Gamma_{_\Phi,_\Psi}=\Gamma_{_A}+\Gamma_{_B}.$ Thus obtained dynamical equations are then used to deduce the ISF as the solutions of these equations in the Fourier domain as discussed in \cite{supp}. For the density component $\Phi$, the ISF is given by,
\be
F_{\Phi}(q,\tau) = S_{\Phi}(q)e^{\tau \beta_{\Phi}(q)},
\label{eqn_isf_phi}
\ee
where $S_{\Phi}(q)$ is defined as the static structure factor which can be obtained from the initial condition and can also be denoted as $F_{\Phi}(q,0)$ and $\tau$ is the time interval. In Fig.\ref{fig:relaxation}b, we plot $F_{\Phi}(q,\tau)/F_{\Phi}(q,0)$, which according to Eq.  (\ref{eqn_isf_phi}) is an exponential function and compare it to the simulation results of the lattice model at a particular value of $q$ where the radial distribution has a maximum. 
We see that the theoretical prediction holds for two orders of magnitude in time before the simulated correlations rapidly decrease. Importantly, active amphiphile turnover does not affect this relaxation dynamics.

The situation is markedly different for the amphiphile component $\oo$. The solution is more involved, as it contains terms stemming from the non-conserved reaction flux because of intermittent dynamics. With a little algebra as shown in \cite{supp}, its ISF reads:

\bea
F_{\oo}(q,\tau) &=& \Bigg( S_{\oo}(q) + \kon - \frac{S_{\Phi \oo}(q)\zeta(q)}{\beta_{\Phi}(q)-\beta_{\oo}(q)} \Bigg) e^{\tau \beta_{\oo}(q)} \nn \\ &+& \Bigg(\frac{S_{\Phi \oo}(q)\zeta(q)}{\beta_{\Phi}(q)-\beta_{\oo}(q)} \Bigg)e^{\tau \beta_{\Phi}(q)}
\label{eqn_int_scattering_om}
\eea
Eq. (\ref{eqn_int_scattering_om}) contains the static structure factor $S_{\oo}(q)$ which involves a single density component and $S_{\Phi \oo}(q)$ coupling both density components as well as terms $\beta_{\Phi}(q),$ $\beta_{\oo}(q,\kon,\kof)$ and $\zeta(q,\kon)$ as discussed in \cite{supp}. We have no direct analytical method to calculate $S_{\oo}(q)$ and $S_{\Phi \oo}(q),$ so we use them as fit parameters when comparing with the simulation results. We have also determined these values numerically in the lattice simulations and found that $S_{\Phi \oo}(q)$ has a negative sign, which indicates that the two density components are anti-correlated. $F_{\oo}(q,\tau)$ is a superposition of two exponentials as revealed in Fig.\ref{fig:relaxation}b.
For $\oo,$ the active system shows an additional fast relaxation time dependent on the amphiphile turnover $\beta_{\oo}(q,\kon,\kof)^{-1}$ and then crosses over to a slower relaxation dynamics common to all components of the system which is independent of turnover $\beta_{\Phi}(q)^{-1}$. 
Lastly, we asked what other signatures of non-equilibrium we could identify by observing the dynamics of the system.

\textit{Local entropy production.-}
Entropy production is a key concept in nonequilibrium statistical physics, quantifying time-reversal symmetry breaking, and thus represents a measure of irreversible changes occurring within the system. Here, we use an information-theoretic approach to quantify the entropy production \cite{nardini} through the methodology rooted in the context of compression algorithms.
Entropy production can be measured in terms of Kullback-Leibler divergence \cite{kl_divergence}, which quantifies the difference between two probability distributions associated with time-forward and time-reversed system's trajectories. It is a positive dimensionless value that signals system's departure from equilibrium.  The methods of measuring local entropy production pioneered in \cite{nardini} and further incorporated the concept of cross-parsing complexity \cite{ziv1,ziv2} formulated as an optimized estimator from the time series of system's dynamics \cite{edgar}.  Following recent work \cite{stefano} that applied this model-free method to a two-dimensional system of active Brownian particles, we use this approach to calculate entropy production in our active microemulsion model (see \cite{supp} for details).

\begin{figure}[ht]
\includegraphics[width=8.6cm,height=2.3cm]{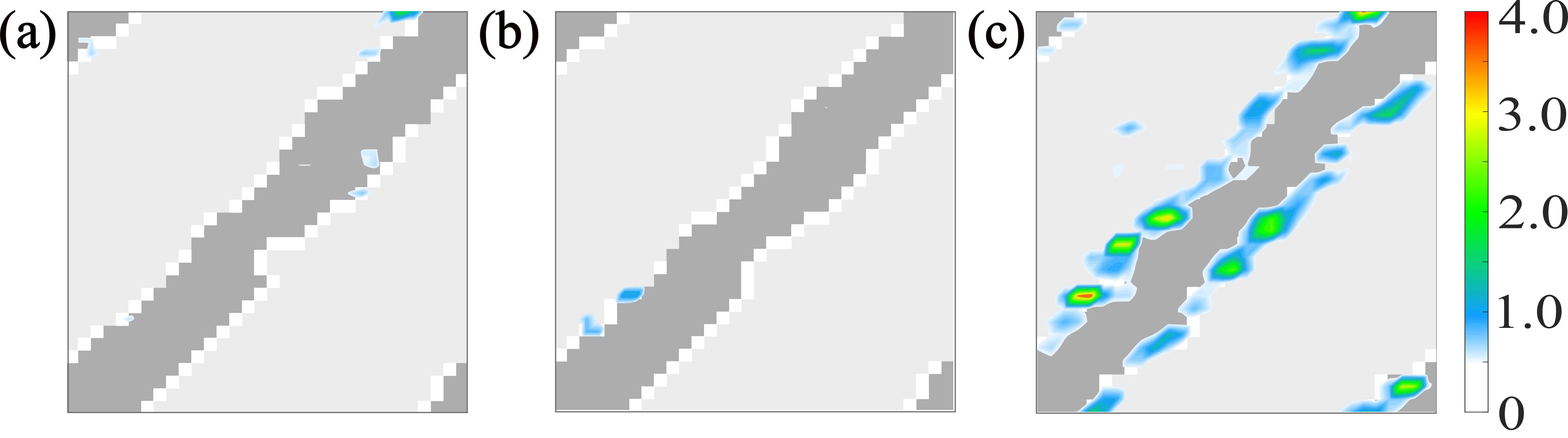}
\caption{Local entropy production for different lattice system setups. (a) Passive system (with no amphiphile turnover). (b) Passive system with amphiphile turnover satisfying thermal equilibrium. (c) Active system with intermittent amphiphile dynamics. Stable $A$-$B$ lamellar interface is generated with a proper choice of model parameters (initial snapshot of the system is shown in greyscale).
High levels of entropy production (color scale) are detected at the $A$-$B$ interface for the active case, whereas no significant entropy production is present for the other two passive cases.  ($A=0.67,$ $B=0.26$, interaction parameters and rates same as in Fig.2.)}
\label{fig:entropy_noneql}
\end{figure}

We start with parameters
leading to stable $A$-$B$ interface (lamellar phase) for both passive and active systems. Furthermore, we explore a third model variant in which amphiphile turnover is allowed, albeit under conditions of thermal equilibrium, (see \cite{supp}). We notice that the $A$-rich and $B$-rich regions have practically no changes of configuration; mostly changes occur at the $A$-$B$ interface. For the passive case this interface, we see no entropy production as indicated with color scale overlaid on top of initial system's configuration (shown in grey scale), Fig.\ref{fig:entropy_noneql}a.
In case of amphiphile turnover under thermal equilibrium, situation is similar (see Fig.\ref{fig:entropy_noneql}b). However, for the nonequilibrium case, we observe a clear entropy production at the interface (see Fig.\ref{fig:entropy_noneql}c) signifying time-irreversible particle interchanges in agreement with previous work \cite{nardini}.

In conclusion, we have shown that the active turnover of amphiphiles in a phase separating system can change both its structure and its dynamics. 
Turnover of amphiphilic activity manifests itself in an additional relaxation time scale emerging in the intermediate scattering function associated with the amphiphile. Consistent with those observations, local entropy production occurs at the interface between the phase separating components. This developed approach could be relevant in the experiments studying chromatin organization. Fluorescent labeling of  RNA and DNA could provide access to structural information \cite{noa} while imaging polymerase molecules could allow to quantify relaxation dynamics.  While the labeling of polymerases can be made state specific (paused vs elongating) \cite{Pancholi}, visualizing the switching of states is challenging; that could be the information required to perform local entropy production measurements. On the side of the future model development, it would be important to investigate the effects of RNA turnover as it happens in the cell nucleus. This work therefore lays grounds for the description of active microphase separated systems as they now recognized across multiple biological processes and active systems.

%%====================================================%%

\pagebreak
\begin{widetext}

\setcounter{equation}{0}
\setcounter{figure}{0}
\setcounter{table}{0}
\setcounter{page}{1}
\setcounter{section}{0}
\makeatletter
\renewcommand{\theequation}{S\arabic{equation}}
\renewcommand{\thefigure}{S\arabic{figure}}
\renewcommand{\thesection}{S\Roman{section}} 
\renewcommand{\bibnumfmt}[1]{[R#1]}
\renewcommand{\citenumfont}[1]{R#1}

\begin{center}\Large{Supplementary Material for ``Non-equilibrium structure and relaxation in active microemulsions"}\end{center}

\subsection{A. Coarse-grained representation}

We start from the microscopic description of the Hamiltonian in terms of the occupancy operator to frame the continuum model via coarse-graining for the microemulsion system in equilibrium. In terms of the occupancy operator $\eta_i^{A,B,\oo,\oi}$ which becomes unity if the site $i$ is occupied by the respective component and zero otherwise, is given by Eq. (1) in main text.
Now, when we introduce the intermittent dynamics $\eta_i^{\oo}  \xrightarrow{\kof}  \eta_i^{\oi} ~,~\eta_i^{\oi} \xrightarrow{\kon}  \eta_i^{\oo}$ are not coupled with system energy, $\oo$'s and $\oi$'s do not appear in the Hamiltonian. They enter later at the dynamical equations. When the intermittent dynamics are in accordance with the system energy, we can write the free energy expression considering the intermittent rates in terms of chemical potential as discussed in sec. D.

Now, by considering the nearest neighbour interactions where the amphiphile-mediated interaction involves three components in a line, we can use the relation $A_iA_{i+1} = (A_i^2 + A_{i+1}^2 - (A_{i+1}-A_i)^2)/2$ and $A_iB_{i+1}=(A_iB_i+A_{i+1}B_{i+1}+A_i(B_{i+1}-B_i)-(A_{i+1}-A_i)B_{i+1})/2,$ 
then the Hamiltonian can be written as,
\bea
{\cal{H}} &=& \frac{1}{2} \sum -J_{AA}\frac{A_i^2+A_{i+1}^2-(A_{i+1}-A_i)^2}{2} - J_{BB}\frac{B_i^2+B_{i+1}^2-(B_{i+1}-B_i)^2}{2} \nonumber \\
&-& - J_{\oo} \oo_{i+1} \Bigg(  A_iB_i + A_{i+2}B_{i+2} + (A_{i+2}-A_i)(B_{i+2}-B_i) \nn \\
&-& 
\frac{A_i^2+A_{i+2}^2-(A_{i+2}-A_i)^2}{2} - \frac{B_i^2+B_{i+2}^2-(B_{i+2}-B_i)^2}{2} \Bigg)
\eea

In the mean-field approximation, $A_i = A+\delta s_i,$ where $A=\la A_i \ra,$ $\delta s_i$ is the corresponding fluctuation and $\la ... \ra$ denotes ensemble average. So, the coarse-grained values of $A_i, B_i, \oo_i$ can be understood as the averages of their microscopic quantities at the position $i.$ The coarse-grained Hamiltonian then reads:

\bea
{\cal{H}} &=& -\frac{J_{AA}}{2} \Big( A^2 - \frac{1}{2} (\nabla A)^2 \Big) -\frac{J_{BB}}{2} \Big( B^2 - \frac{1}{2} (\nabla B)^2 \Big) \nonumber \\ 
&-& J_{\oo} \oo \Big( 2AB - 4\nabla A \cdot \nabla B - (A^2 + B^2 -2(\nabla A)^2 - 2(\nabla B)^2) \Big)
\eea

The corresponding free energy density is shown by Eq. (2) in the main text, here we have considered the same order of expansion for all the components, \textit{i.e.} up to two lattice spacing. 

Eq. (2) in the main text consists of the interaction energy terms as well as the terms due to free energy of mixing. The free energy of mixing is a concept in thermodynamics that describes the increase in entropy when two or more different substances are mixed together. The mixing always increases the entropy of the system, consistent with the second law of thermodynamics. It is worth noting that the expression assumes an ideal mixing scenario, where the interactions between different components are negligible compared to the thermal energy of the system. In more complex scenarios, such as non-ideal solutions or mixtures with significant inter-particle interactions, additional terms may need to be included in the free energy expression. The entropy of random mixing is written as a function of the volume fraction of each components as shown by Eq. (2) in main text. Here, we have only considered attractive interaction between similar components e.g. $J_{AA}$ and $J_{BB}$, but $J_{AB}=0.$

Now, we formulate the dynamics that governs the microphase separation. The equations that incorporate density components $A$ and $B$ are applicable in both equilibrium and non-equilibrium conditions. However, the equation that includes $\oo$ is exclusively pertinent to nonequilibrium conditions due to its association with intermittent dynamics not coupled to thermodynamic equilibrium. Conservation of components $A,B,\oo$ and $\oi$ can be represented by the continuity equations. In the linear response regime, the dynamical equations are given by Eq. (3-4) in the main text. For each component in the system they can be written explicitly as,
\bea
\partial_t A &=& \Gamma_A\nabla \cdot \nabla \Bigg( -J_{AA} \Big(A + \frac{\nabla^2A}{2} \Big)
- J_{\oo} \Big(2\oo B + 4\oo \nabla^2 B -2\oo A -4\oo \nabla^2 A\Big)
+ \logn A -\logn \Big(1-A-B-\oo \Big)\Bigg) \nonumber \\
\partial_t B &=& \Gamma_B \nabla \cdot \nabla \Bigg( -J_{BB} \Big(B + \frac{\nabla^2B}{2} \Big)
- J_{\oo} \Big(2\oo A + 4\oo \nabla^2 A -2\oo B -4\oo \nabla^2 B \Big)
+ \logn B -\logn \Big(1-A-B-\oo \Big)\Bigg) \nonumber \\
\partial_t \oo &=& \Gamma_{\oo} \nabla \cdot \nabla \Bigg( -J_{\oo}\Big(2AB+4 \nabla A \cdot \nabla B - A^2-B^2+2(\nabla A)^2 +2(\nabla B)^2\Big) + \logn \oo -\logn \Big(1-A-B-\oo \Big) \Bigg) \nn \\
& &  -\kof \oo +\kon (1-A-B-\oo)
\label{eqn_dynamical_full}
\eea

For the sake of simplifying the equations, we have introduced two modified local densities $\Phi=A+B$ and $\Psi=A-B.$ The dynamical equations with the modified local densities can be written as:
\bea
\partial_t \Phi &=& \Gamma_{\Phi} \nabla \cdot \nabla \Bigg(-\frac{J_{AA}}{2}(\Phi + \nabla^2 \Phi) + \frac{J_{AA}}{4} \nabla^4 \Phi + \frac{1}{2}\logn \frac{\Phi+\Psi}{2} + \frac{1}{2}\logn \frac{\Phi-\Psi}{2} - \logn (1-\Phi) \Bigg) \nonumber \\
\partial_t \oo &=& \Gamma_{\oo} \nabla \cdot \nabla \Bigg( -J_{\oo}\Big(-\Psi^2 +2 (\nabla \Psi)^2\Big)+\logn \oo -\logn (1-\Phi-\oo ) \Bigg) - \kof \oo +\kon (1-\Phi-\oo)
\label{eqn_dynamical_full_phi_psi_om}
\eea

Here, the diffusion coefficient, represented by $\Gamma_{_{A,B,\oo}}$, corresponds to the movement of particles at unit temperature. If the temperature is held constant and it is assumed that all particles are transitioning through nearest neighbor exchange at an equivalent rate, then the diffusion coefficients for all particle varieties are deemed identical. The diffusion coefficient for the modified local densities turn out to be $\Gamma_{_\Phi,_\Psi}=\Gamma_{_A}+\Gamma_{_B}.$

We next take the two density components $\Phi$ and $\oo$ to investigate the relaxation dynamics. For this, we need to derive the time evolution of the density-density correlation, \textit{i.e.} the Intermediate Scattering Function (ISF). As cross-correlation functions are easier to analyse in Fourier domain, we do so for the rest of the derivation. 

We start with a small perturbation from the steady state $\Phim$ and $\oom$ by setting $\Phi=\Phim + \Phit$ and $\oo=\oom + \oomt.$ Here $\Phit$ and $\oomt$ denote small fluctuations. Using linear stability analysis in Fourier domain, Eq. (\ref{eqn_dynamical_full_phi_psi_om}) becomes:

\bea
\partial_t \Phit &=& \Gamma_{\Phi} \Bigg(-\Big(\frac{1}{\Phim(1-\Phim)}\Big)q^2  + \frac{J_{AA}}{2} (q^2 -q^4) - \frac{J_{AA}}{4} q^6  \Bigg)\Phit \nonumber \\
\partial_t \oomt &=& - \Gamma_{\oo}\Big(\frac{1}{\oom} + \frac{1}{1-\Phim - \oom}\Big)q^2 \oomt - \Gamma_{\oo} \Big( \frac{1}{1-\Phim-\oom}\Big) q^2\Phit - \kof \oomt +\kon(1-\Phit - \oomt)
\label{eqn_lin_stability}
\eea

Eq. (\ref{eqn_lin_stability}) can be rewritten in a compact form as,

\bea
\partial_t \Phit &=& \beta_{\Phi}(q) \Phit  
\label{eqn_dynamic_compact_1} \\
\partial_t \oomt &=& \beta_{\oo}(q)\oomt + \zeta (q)\Phit + \kon
\label{eqn_dynamic_compact_2}
\eea

where we have defined
\bea
\beta_{\Phi}(q) &=&\Gamma_{\Phi}\frac{J_{AA}}{2}(q^2 - q^4) - \Gamma_{\Phi} \frac{J_{AA}}{4}  q^6 -\Gamma_{\Phi}\Big(\frac{1}{\Phim(1-\Phim)}\Big)q^2 \nn, \\
\beta_{\oo}(q) &=& -\Gamma_{\oo}\Big( \frac{1}{\oom} + \frac{1}{1-\Phim -\oom} \Big) q^2 -\kon-\kof \nn, \\
\zeta(q) &=& -\Gamma_{\oo} \Big(\frac{q^2}{1-\Phim -\oom}\Big) - \kon. \nn
\eea

\subsection{B. Structure factor analysis}
{One thing worth mentioning here is that, we ensured all measurements were taken after a sufficiently long equilibration period. For reference, we compared the structure factor at the final time step of the recording with that at the midpoint of the total time span. Our results indicate that the peak of the structure factor remains at the same position for this two cases in both active and passive states, and the profiles are almost identical, as shown in Fig. \ref{compare}.}

\begin{figure}[ht]
\subfloat{{\includegraphics[width=6cm]{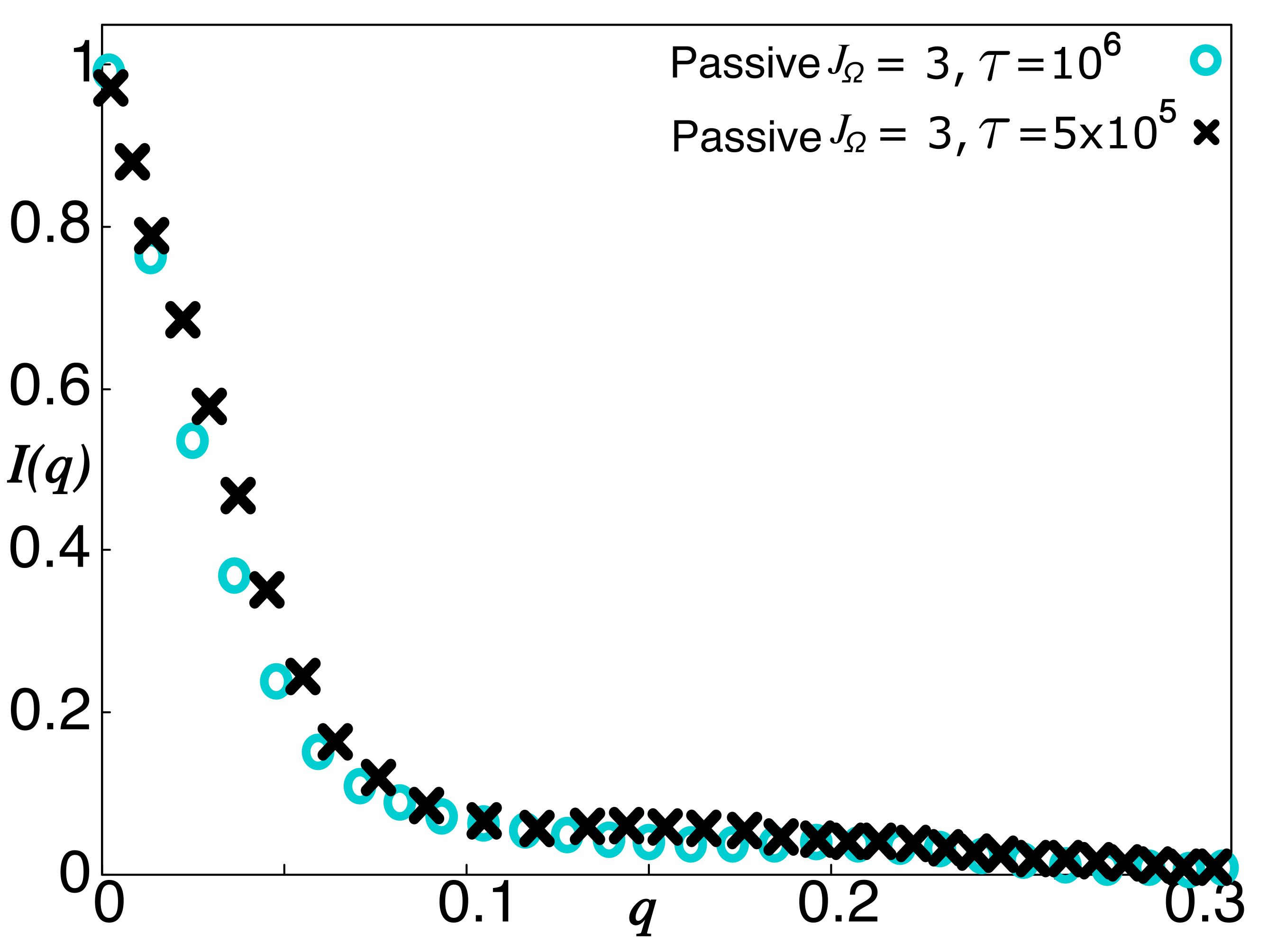} }}
\subfloat{{\includegraphics[width=6cm]{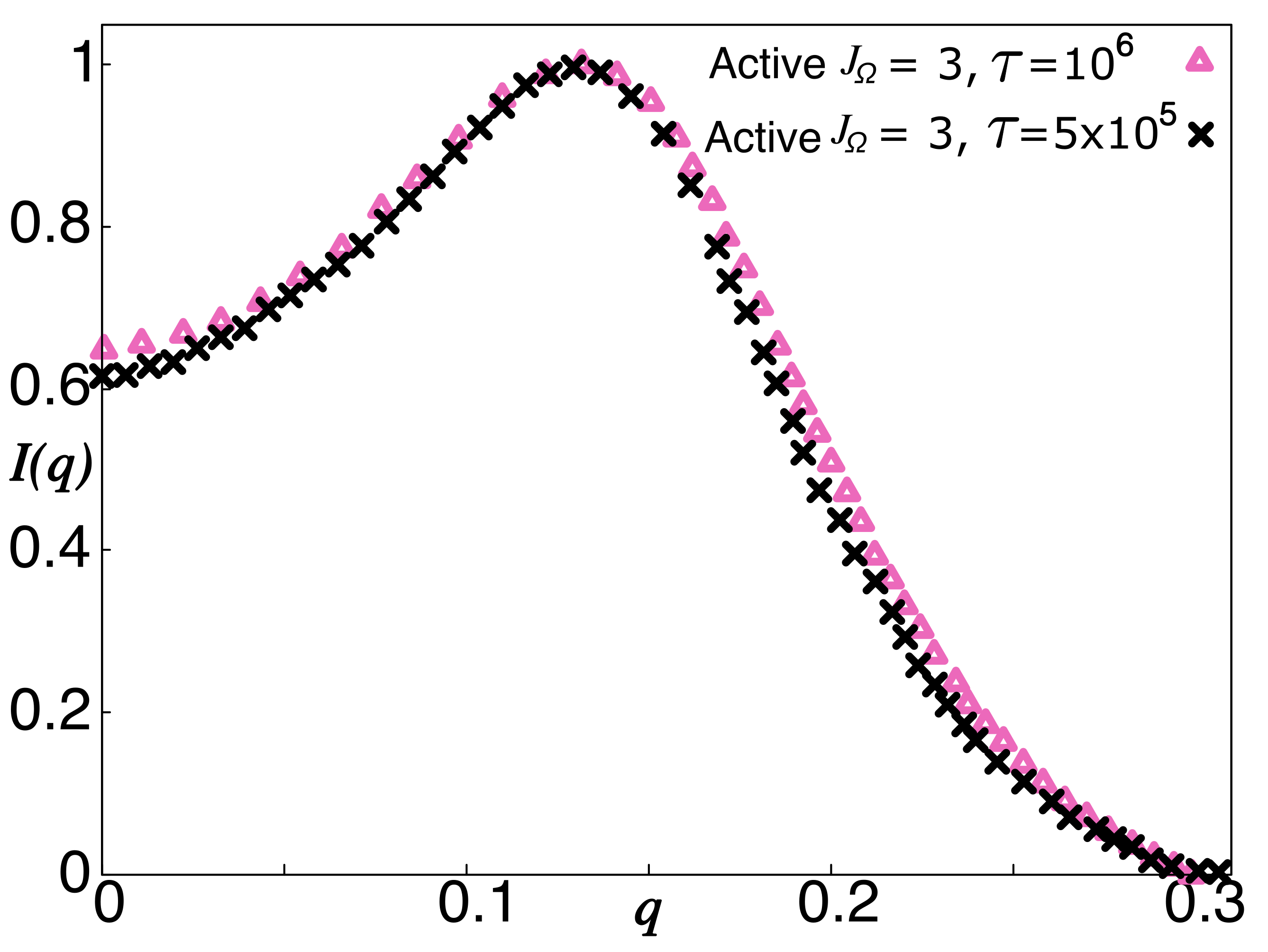} }}
\caption{Comparison of the structure factor at the final time step with that at half of the total time span for both passive and active systems for $A = B = 0.35,$ $\Omega_{\text{Total}} = 0.30,$ $J_{AA} = J_{BB} = 2.5$, $\kon = 0.90$, $\kof = 0.10$ and $J_{\Omega}=3.0.$ Here we wanted to look at the evolution of structures over time. The figure clearly shows that by the timestep $\tau=5 \times 10^5,$ the structure has stabilized, and no further coarsening takes place.}
\label{compare}
\end{figure}

\begin{figure}[ht]
\includegraphics[width=6cm]{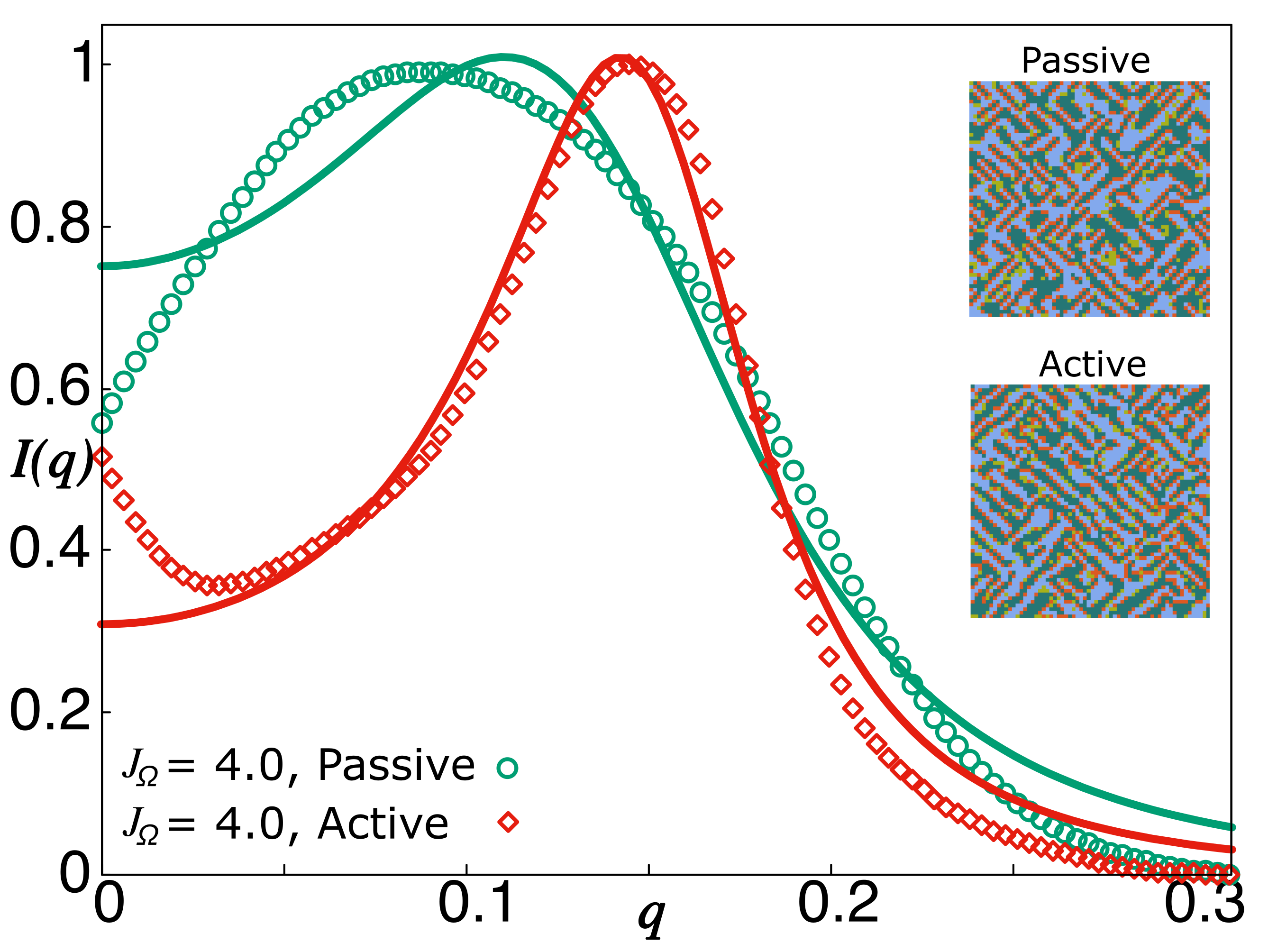}
\caption{Radial intensity distribution in Fourier domain of lattice configuration in passive and active systems for $A = B = 0.35,$ $\Omega_{\text{Total}} = 0.30,$ $J_{AA} = J_{BB} = 2.8$, $\kon = 0.90$, $\kof = 0.10$ and $J_{\Omega}=4.0$. The distribution is fitted with the phenomenological Landau theory of microemulsion ($Y_{TS}(q)$). It indicates that the phenomenological theory fits better for the active structure. The inset show the respective lattice structures.}
\label{peak}
\end{figure}

{The expression for the structure factor $Y_{TS}(q)$ from a phenomenological equilibrium microemulsion theory \cite{teubner} can be utilized as a fitting function for active microemulsion: $Y_{TS}(q) \sim (a+c_1 q^2 + c_2 q^4)^{-1}.$ With a negative $c_1$, this expression initially peaks before decreasing as $q^{-4}$ as shown in Fig. \ref{peak}.}

\subsection{C. Intermediate scattering function}

In the realm of statistical physics, particularly in the study of systems undergoing dynamic processes, the intermediate scattering function serves as a fundamental quantity. It plays a pivotal role in characterizing the temporal evolution of correlation functions in these systems. The ISF is primarily utilized in the context of dynamic structure factor, which describe the spatial and temporal correlations of fluctuations in a system's density or other relevant order parameters. Specifically, the ISF quantifies how these correlations evolve over time.
Technically, the intermediate scattering function $F(q,t)$ is defined as the time correlation function of the fluctuations in the scattered intensity at a given wave vector $q.$

In simulations, the ISF is obtained by measuring the time-dependent structure factor, which captures the spatial correlations of fluctuations at different times. By analyzing the decay of the ISF over time, one can extract crucial information about the dynamic behavior of the system, including the presence of relaxation processes and the nature of collective phenomena such as diffusion, it also sheds light on the dynamic behavior of matter at various length and time scales.

As we already have the dynamical equations in the Fourier space given by Eq. (\ref{eqn_dynamic_compact_1}) and Eq. (\ref{eqn_dynamic_compact_2}), we can proceed to get the ISF from there. The solution for Eq. (\ref{eqn_dynamic_compact_1}) is simple and given by Eq. (5) in the main text. The dynamical structure factor can be obtained by the Laplace transformation of Eq.  (\ref{eqn_dynamic_compact_1}) and Eq.  (\ref{eqn_dynamic_compact_2}). The Laplace transform of the time derivative in general yields ${\cal{L}}(f'(t))=\omega f(\omega)-f(0).$ Thus from Eq.  (\ref{eqn_dynamic_compact_1}) and Eq.  (\ref{eqn_dynamic_compact_2}) we get,

\bea
w \Phit -\Phit(0) &=& \Gamma_{\Phi} \Bigg(-\Big(\frac{1}{\Phim(1-\Phim)}\Big)q^2 \Phit + \frac{J_{AA}}{2} (q^2 -q^4)\Phit - \frac{J_{AA}}{4} q^6 \Phit \Bigg) \nn \\
&=& \beta_{\Phi}(q)\Phit 
\label{eqn_fourier_1}
\\
w \oomt - \oomt (0) &=& \Gamma_{\oo} \Bigg(-\Big(\frac{1}{\oom} + \frac{1}{1-\Phim - \oom}\Big)q^2 \oomt - \Big( \frac{1}{1-\Phim-\oom}\Big) q^2\Phit \Bigg) - \kof \oomt +\kon(1-\Phit - \oomt) \nn \\
&=& \beta_{\oo}(q) \oomt + \zeta(q)\Phit + \kon
\label{eqn_fourier_2}
\eea

The dynamical structure factor from Eq. (\ref{eqn_fourier_1}) and Eq. (\ref{eqn_fourier_2})
can be written as,

\bea
S_{\Phi}(q,w) &=& \frac{S_{\Phi}(q)}{w-\beta_{\Phi}(q)} \label{eqn_dyn_sf1} \\
S_{\oo}(q,w) &=& \frac{S_{\oo}(q)}{w-\beta_{\oo}(q)} + \frac{\zeta(q)S_{\Phi \oo}(q)}{\Big(w-\beta_{\Phi}(q)\Big)\Big( w-\beta_{\oo}(q)   \Big)}  + \frac{\kon}{w-\beta_{\oo}(q)}
\label{eqn_dyn_sf2}
\eea
where $S_{\Phi}(q)$ and $S_{\oo}(q)$ are the static structure factor for the respective density components $\Phi$ and $\oo$ respectively. 
$S_{\Phi \oo}(q)$ is the coupled static structure factor and 
$S_{\Phi \oo}(q,w) = \frac{S_{\Phi \oo}(q)}{w-\beta_{\Phi}(q)}$ is the coupled dynamical structure factor involving both density components $\Phi$ and $\oo.$ These are defined mathematically as:
\bea
S_{\Phi}(q,w) &=& \int_0^{\infty} e^{w\tau} \Phi(q,0)\Phi^*(q,\tau)d\tau \label{eqn_dyn_str_factor1} \\
S_{\Phi\oo}(q,w) &=& \int_0^{\infty} e^{w\tau} \Phi(q,0)\oo^*(q,\tau)d\tau
\label{eqn_dyn_str_factor2}
\eea
There is no direct method to calculate the static structure factors $S_{\oo}(q)$ and $S_{\Phi \oo}(q)$ through coarse-grained analytical treatment. To this end, we measured these values numerically and noticed that $S_{\Phi \oo}(q)$ has negative sign, which indicates the two density components $\Phi$ and $\oo$ are anti-correlated. We have assigned these terms as fitting parameters in the analytical expression to get the final solution and compare with the numerical result. 

Now, the solution of the Eq. (\ref{eqn_dynamic_compact_2}) is not so straight-forward to find out the ISF. So, we adopted another way to deduce the ISF through inverse Laplace transformation of the dynamical structure factor from Eq. (\ref{eqn_dyn_sf2}). In this way we get the ISF for the density component $\oo$ as shown by Eq.  (6) in the main text. 

To measure the relaxation in terms of the ISF numerically, we start with the time evolution of lattice configurations from a random initial state. We generate time evolution of lattice configuration images which consist of lattice sites occupied with either of the density component $A,$ $B$ or $\oo$ only for which the relaxation is measured. For example, to measure the relaxation of $A,$ each image only consists of sites with $A$ and rest of the lattice space is empty. Similarly, to measure relaxation of $\Phi=A+B,$ each image has sites only occupied with $A$ and $B.$ To suppress the noise, we measure the average occupancy of each lattice site considering all the time evolution configurations, and subtract it from every lattice site of each configuration. Then we transform each lattice configuration image in the Fourier domain through the two dimensional Fast Fourier Transform (FFT2).

In our simulations, ISF is obtained by measuring the density correlation of fluctuations at different times as,
\be
F(q,\tau) = \frac{\sum_{t=1}^{M-\tau } \rho(q,t)\rho^*(q,t+\tau)}{M-\tau}
\label{eqn_isf_corr}
\ee
where $\rho(q,t)$ is the density map of an image at time $t$ in Fourier domain, $\tau$ is the time interval over which the correlation is measured between two images and $M$ steps are used for time averaging. For the $q-$value we chose the particular wave number in radial distribution of the images for which the intensity is maximum.

The density component $\Phi$ behaves identically for active and passive case and shows an exponential decay as given by Eq. (5) in the main text. Whereas the density component $\oo$ behaves differently in terms of relaxation for passive and active cases. As in the active case it involves intermittent dynamics with two rates $\kon$ and $\kof$ as given by Eq. (6) in the main text. For active case it turns out to be a superposition of two exponential functions.
This is why we observed the relaxation of $F_{\oo}(q,\tau)$ as a rapid decay followed by a slower one.

\subsection{D. Local entropy production}

Local entropy production ($\ep$) is a concept within the framework of irreversible thermodynamics that quantifies the rate at which entropy increases within a small region of a system. It provides valuable insights into the irreversible processes occurring at the microscopic level. $\ep$ focuses on analyzing entropy changes within small spatial regions of the system.
To measure local $\ep$, we track the system configurations for a given observation interval. The main idea is to observe the events of time reversal symmetry breaking (TSRB) at the interface of $A$ and $B$ components separated by $\oo$ where the reversibility in lattice configuration is not preserved due to the system dynamics. 
We have measure the $\ep$ originating due to local microscale processes e.g. particle diffusion, energy transfer, or other irreversible interactions between neighboring lattice sites.

$\ep$ can be quantified by the Kullback-Leibler divergence \cite{kl_divergence} which is defined as the difference between two probability distributions (PDF). Rather than measuring PDF of local configurations, it is a standard procedure to use the method based on the proposal first illustrated in terms of compression algorithms but without any association with the physical entropy \cite{ziv1,ziv2}. In this methodology, the system configurations in each time steps of a certain interval, are reduced to an array of numbers.  A quantity called cross parsing complexity ${\cal{C}}$ which is defined as the sum of sequentially drawn longest arrays of numbers from one sequence while parsing through or iterates through all elements of another sequence is coined.  One sequence is derived from the randomly drawn system configuration, reduced into number of arrays for each forward time-step. The other sequence contemplates the time-span in reverse direction.  Here ${\cal{C}}$ is defined as the difference between the cross parsing involving forward and reverse sequence, and the self incremental parsing of the forward sequence.  Later,  a large error due to mixing of two types of parsing namely cross parsing and sequential self incremental parsing in this method is reported and a corrected $\ep$ estimator by evaluating the cross parsing between different segments of the same trajectory and then subtracted that from the earlier estimator is proposed \cite{edgar}.  We have used the corrected $\ep$ estimator and followed the model free method for our lattice system to extract $\ep$ from system dynamics.

To do that, we start by overlaying a small grid of size $2 \times 1$ over the system and record the lattice configuration in each time-step. For a particular position of the grid, one can have a set of configurations corresponding to each time step in forward time-span. From this set we can get the quantity ${\cal{Z}}$ by random sequentially drawn $N$ number of configurations to form a forward sequence. Similarly, one can get a reverse sequence from the set created by having the configurations for each time step in the backward time-span denoted by ${\cal{Z}}^*.$ Considering these two sequences, we can get the $\ep$ estimator proposed by the earlier method \cite{ziv1}.  Now, to evaluate the correction term  suggested by the later article \cite{edgar}, we divide the forward sequence into two equal parts, and measured the cross parsing between them. This term is now subtracted from the earlier estimation to get the correct $\ep$ estimator, which can be written as:

\bea
{\cal{E}}_p &=& \lim_{N\to\infty} \frac{1}{N} \Big( {\cal{C}}({\cal{Z}}||{\cal{Z}}^*)\Big) - \frac{1}{N/2}\Big({\cal{C}}({\cal{Z}}_{N/2}^N||{\cal{Z}}_1^{N/2}) \Big)
\label{eqn_cpc1} \\
&=& \lim_{N\to\infty} \frac{1}{N}\Big({\cal{C}}({\cal{Z}}||{\cal{Z}}^*) \ln N - {\cal{C}}({\cal{Z}})\ln{{\cal{C}}({\cal{Z}}})\Big) - \frac{1}{N/2}\Big({\cal{C}}({\cal{Z}}_{N/2}^N||{\cal{Z}}_1^{N/2}) \ln (N/2) - {\cal{C}}({\cal{Z}}_{N/2}^N) \ln{{\cal{C}}({\cal{Z}}_{N/2}^N)} \Big)
\label{eqn_cpc2}
\eea

The above equation consists of two cross-parsing complexities and specifically, the second term of Eq. (\ref{eqn_cpc1}) represents the correction term proposed by \cite{edgar}. While in Eq. (\ref{eqn_cpc2}) the second term from each cross-parsing  corresponds to the Shannon entropy of the constituent sequence.
In Fig.4 of the main text, we have shown the entropy production-maps of the lattice system measured in this way for three different scenarios namely the passive case which is equilibrium and has no intermittent amphiphile dynamics, then the case with intermittent amphiphile dynamics but within chemical equilibrium, and, finally, the active case with intermittent amphiphile dynamics which drives the system out of equilibrium.

The main objective is to extract meaningful information or patterns from heterogeneous or diverse data sets that might be structured differently or follow varying conventions. It can be noticed from the entropy production-map that for passive case, the local $\ep$ is zero everywhere in the lattice. It could also possible that they become very low approaching towards zero or only limited to positions where a few irreversible particle interchanges occur, which we attribute to numerical fluctuations. Intermittent dynamics within chemical equilibrium is showing also no $\ep$. However, for the active case a high $\ep$ is observed throughout the interface. So, here we have shown that it is possible to identify the regions of the system that are driven out of equilibrium through this information-theoretic approach and this is also a promising tool to define such regions in experimental data.

It is also important to note that, we have introduced a scenario where the intermittent dynamics is consistent with chemical equilibrium to see whether the $\ep$ is mostly contributed from the intermittent dynamics itself or from the irreversible nonequilibrium type of dynamics. What we found is that, in the second case, though there are intermittent dynamics, but still no $\ep$ as the system is at equilibrium. Now we will discuss the case of intermittent dynamics within chemical equilibrium through mean-field analysis.

\subsection{E. Dynamics within chemical equilibrium}
\label{subsec_D}

We start with the system where the intermittent activity is within the thermodynamic equilibrium. The detailed balance condition for thermodynamic equilibrium is given by,
\bea
\oo \kof = \oi \kon
\label{cond_ceq_1}
\eea

The free energy expression as Eq. (2) in the main text does not involve chemical reaction or intermittent dynamics.  As molecular energy transformation is responsible for chemical reaction rates at the most fundamental level, chemical kinetics must provide some  characteristics of molecular energies. So, a reversible chemical reaction involving amphiphilic-inert transition, gives rise to an additional term consisting of transition rates within the free energy density. Let's denote $\mu_{\oo}$ and $\mu_{\oi}$ as the chemical potentials of amphiphilic and inert particles respectively. The chemical potential represents the change in the free energy of the system when the number of particles of a particular species changes.

Now, the chemical potential of amphiphilic and inert particles can be written in terms of energy unit $k_BT$ as
$\mu_{\oo} = -\logn (\oo N)$ and
$\mu_{\oi} = -\logn (\oi N),$
where $N$ is the total number of particles in the system, then $\oo N$ and $\oi N$ represents the total number of amphiphilic and inert particles respectively. As there is no empty space in the system, so $N=L \times L.$ So, we can define the Free energy in thermodynamic equilibrium as,
\bea
{\cal{F}}' =  \oo N \mu_{\oo} +  \oi N \mu_{\oi}
\label{cond_ceq_2}
\eea

So, adding this term with the free energy of interaction and mixing in the system, we get,

\begin{widetext}
\bea
{\cal{F}}^{CEq} &=&-\frac{J_{AA}}{2} \Big( A^2 - \frac{1}{2}(\nabla A)^2 \Big)-\frac{J_{BB}}{2}\Big( B^2 - \frac{1}{2}(\nabla B)^2 \Big) - J_{\oo} \Big( \oo (2AB-4 \nabla A . \nabla B)) \nonumber \\
&-& \oo(A^2+B^2-2(\nabla A)^2 -2(\nabla B)^2 \Big) +  \Big( A\logn A + B\logn B + \oo \logn \oo + (1-A-B-\oo) \logn (1-A-B-\oo) \Big) \nonumber \\
&-& \oo N \logn( \oo N) - (1-A-B-\oo) N \logn \big((1-A-B-\oo)N\big)
\label{free_en_cont_onoff_eql}
\eea
\end{widetext}

It can be noticed that the last two terms of Eq.  (\ref{free_en_cont_onoff_eql}) are in accordance with the intermittent dynamics within thermodynamic equilibrium. Then the dynamical equation for $\oo$ can be written as,
 
\begin{widetext}
\bea
 \partial_t \oo^{CEq} &=& \Gamma_{\oo} \nabla \cdot \nabla \Bigg( -J_{\oo}\Big(2AB+4 \nabla A . \nabla B - A^2-B^2+2(\nabla A)^2 +2(\nabla B)^2\Big) \nonumber \\
 &-& (N-1) \Big(\logn \oo - \logn (1-A-B-\oo) \Big) \Bigg)
\label{eqn_dynamical_omega_onoff_eql}
\eea
\end{widetext}

Adopting similar methodology with the modified local densities $\Phi$ and $\oo$ and then applying small perturbation from the steady state in the Fourier space, Eq.  (\ref{eqn_dynamical_omega_onoff_eql}) yields,

\begin{widetext}
\bea
\partial_t \oomt^{CEq} &=&  \Gamma_{\oo}(N-1)\Big(\frac{1}{\oom} + \frac{1}{1-\Phim - \oom}\Big)q^2 \oomt + \Gamma_{\oo} (N-1)\Big( \frac{1}{1-\Phim-\oom}\Big) q^2\Phit 
\label{eqn_dynamical_omega_onoff_eql_lin}
\eea
\end{widetext}

Eq. (\ref{eqn_dynamical_omega_onoff_eql_lin}) can be written in a compact form as,

\bea
\partial_t \oomt^{CEq} &=& \beta^{CEq}_{\oo}(q) \oomt  + \beta^{CEq}_{\Phi}(q) \Phit
\label{eqn_onoff_compact2}
\eea

Where we have defined
\bea
\beta^{CEq}_{\oo}(q) &=&\Gamma_{\oo} (N-1) \Big( \frac{1}{\oom} + \frac{1}{1-\Phim-\oom}\Big) q^2 \nn \\
\beta^{CEq}_{\Phi}(q) &=& \Gamma_{\oo} (N-1)\Big(\frac{1}{1-\Phim-\oom}\Big)q^2 \nn
\eea

Now, here we notice that the expression Eq. (\ref{eqn_onoff_compact2}) is independent of $\kon$ and $\kof,$ and in that sense it is different from the active system where the  dynamical equation involving $\oo$ depends on the intermittent rates that drive the system out of equilibrium.

\end{widetext}
\end{document}